\documentclass[prx,twocolumn]{revtex4-2}
\usepackage{graphicx}
\usepackage{amsmath}
\usepackage{natbib}
\usepackage{epstopdf}
\usepackage{xcolor}
\newcommand{\n}{\nonumber}
\newcommand{\bn}{\begin{eqnarray}}
\newcommand{\en}{\end{eqnarray}}
\newcommand{\eml}{\end{multline}}
\newcommand{\bml}{\begin{multline}}
\newcommand{\h}{\hspace}

\newcommand{\op}[1]{\hat{#1}}
\newcommand{\bra}[1]{\langle{#1}|}
\newcommand{\ket}[1]{|{#1}\rangle}

\begin{document}

\title{Spin Orbit and Hyperfine Simulations with Two-Species Ultracold Atoms in a Ring}
 \author{Allison Brattley $^1$ Tom\'a\v{s} Opatrn\'y $^2$ and Kunal K. Das $^{3,4}$}
  \affiliation{$^1$ Department of Physics, Massachusetts Institute of Technology, Cambridge, MA 02139, USA}
  \affiliation{$^2$ Department of Optics, Palacký University, 771 46 Olomouc, Czech Republic}
 \affiliation{$^3$ Department of Physical Sciences, Kutztown University of Pennsylvania, Kutztown, Pennsylvania 19530, USA}
  \affiliation{$^4$ Department of Physics and Astronomy, Stony Brook University, New York 11794-3800, USA}

\begin{abstract}
    A collective spin model is used to describe two species of mutually interacting ultracold bosonic atoms confined to a toroidal trap. The system is modeled by a Hamiltonian that can be split into two components, a linear part and a quadratic part, which may be controlled independently. We show the linear component is an analog of a Zeeman Hamiltonian, and the quadratic component presents a macroscopic simulator for spin-orbit and hyperfine interactions. We determine a complete set of commuting observables for both the linear and quadratic Hamiltonians, and derive analytical expressions for their respective spectra and density of states. We determine the conditions for generating maximal entanglement between the two species of atoms with a view to applications involving quantum correlations among spin degrees of freedom, such as in the area of quantum information.
\end{abstract}

\maketitle

\section{Introduction}
The intrinsic spin of particles along with quantized orbital angular momentum have been a defining part of quantum mechanics since the Stern-Gerlach experiment \cite{Cohen_Tannoudji}. The angular momentum of atoms has been central to the progress of atomic, molecular and optical physics \cite{Mandel-Wolf,Metcalf-book}. In recent decades, the spin degree of freedom of electrons has broadened traditional electronics to encompass the growth field of spintronics \cite{Spintronics-RMP}. Manipulation and utility of spin degrees of freedom rely upon various interactions involving spin and orbital angular momentum, be it  mutual interactions of electron spins, spin-orbit coupling or coupling involving nuclear spins. Traditionally, those interactions have been tied to the size and time scales of individual atoms and molecules. It would be desirable to access the same physics at larger and slower scales, both for the purpose of understanding the dynamics in real time and for broadening the scope of applications. We propose a way to simulate general interactions of spin and angular momentum with translational degrees of freedom with two species of Bose-Einstein condensate (BEC) in toroidal traps.

We have modeled such a system in a recent paper \cite{Opatrny-Das-two-species}, and showed that it can be described in terms of collective spin operators \cite{Kitagawa}; with a Hamiltonian comprising of terms linear and quadratic in angular momentum operators for the two species. We showed that the interactions manifest in the collective spin operators in the quadratic component lead to strong entanglement. In order to further our understanding, we require a complete set of commuting observables (CSCO) for the relevant quadratic Hamiltonian, which we determine in this paper, along with analytical expressions for its spectrum. Then we demonstrate how this system can be adapted to simulate both spin-orbit and hyperfine interactions with collective spin operators.

\begin{figure}[t]
\centering
%\vspace{-0.2\linewidth}
\includegraphics[width=\columnwidth]{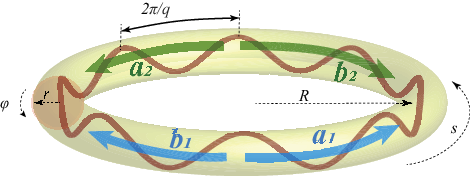}
\caption{Schematic of physical model. Two species of ultracold atoms ($i=1,2$) are confined to a toroidal trap with an azimuthal lattice potential of period $2\pi/q$. The two lowest counterpropogating modes for either species are denoted by $a,b$; and cylindrical coordinates by $(s,r,\phi)$.}
\label{ring_schematic}
\end{figure}

Interactions between collective spins also allow the modeling of quantum correlations associated with multiparticle entanglement, so the ring system can serve to model some of the most intriguing aspects of quantum mechanics, such as EPR and Bell's inequalities. Highly entangled nonclassical states of collective atomic spins can have applications in metrology \cite{Oberthaler-Nature,Riedel-Nature} and for quantum computation \cite{Opatrny-PRL}. A significant advantage of the rings with ultracold atoms is that it offers a macroscopic system, where the degrees of freedom involved are the collective states of the atoms rather than electronic states. This offers a route to a more robust platform for quantum technologies while also allowing for inclusion of continuous variable quantum operations \cite{Lloyd-Braunstein-continous-variable}. Anticipating such applications, we analyze the conditions for maximizing the degree of entanglement and derive analytical expressions for the density of states for relevant limiting cases.

A ring trap with  ultracold atoms has led to several interesting experiments with both bosonic atoms \cite{ramanathan,Phillips_Campbell_superfluid_2013, Campbell_resistive-flow,Phillips_Campbell_hysteresis,Campbell-current-phase, Hadzibabic-2012, Hadzibabic-spinor,von_Klitzing_BEC_accelerator} and fermionic atoms \cite{Kevin_Wright_PRL, Wright_ring2023}. The azimuthally unbounded configuration is a natural choice for studying superfluidity and persistent flow. Such ring systems are developed in laboratories as a platform for diverse applications such as atomtronics circuits \cite{ramanathan, Phillips_Campbell_hysteresis, Campbell_resistive-flow} and  miniature accelerator rings \cite{von_Klitzing_BEC_accelerator}. With the inclusion of an azimuthal lattice, the ring system has been shown to be a viable quantum platform to study a broad range of phenomena: nonlinear physics in a closed system \cite{Das-Huang, Das-Tekverk-Siebor, Opatrny-Kolar-Das-rotation}, spin-squeezing \cite{Opatrny-Kolar-Das-LMG}, physics tied to the quantum Hall effect \cite{Das-Christ, Das-PRL-localization,Das-ring-real-space} and as a simulator for quantum optics with the circulating modes simulating electrons within atoms \cite{Das-Brooks-Brattley}. Two species of atoms in a ring-shaped trap have already been demonstrated in experiments \cite{Hadzibabic-spinor}.

In Sec.~II, we derive the representation of our physical model in terms of collective spin operators. We then present the CSCO for both the linear and the quadratic parts of the Hamiltonian in Sec.~III and derive analytical expressions for their spectrum. In Sec.~IV, we show how this system can be adapted to simulate spin-orbit and hyperfine interactions, illustrated with relevant examples. We analyze the conditions for achieving maximum entanglement in Sec.~V and then derive analytical expressions for the density of states in Sec.~VI, before concluding with a summary of our results and outlook.

\section{Collective Spin Hamiltonian}

The details of our physical model are presented in a recent paper \cite{Opatrny-Das-two-species}; here we will only present the salient points. We consider two species of Bose-Einstein condensate (BEC), labelled $j=1,2$ in a toroidal trap. Taking the minor radius $r$ to be much smaller than the major radius $R$ it can be treated as a cylinder ${\bf r}=(s,r,\phi)$ with periodic boundary conditions on $s$. Assuming strong confinement, transverse degrees of freedom $(r,\phi)$ are integrated out to yield an effective one dimensional (1D) Hamiltonian
\begin{eqnarray}
\op{H}&=&\int_0^{2\pi R}\h{-3mm}{\rm d}s\left[{\small\sum_{i=1,2}}\op{\Psi}_i^\dagger
{\left(- \frac{\hbar^2}{2m_i}\partial^2_s+U_i+\frac{g_i}{4\pi l_i^2} \op{\Psi}_i^\dagger\op{\Psi}_i\right)\op{\Psi}_i}\right.\n\\
&&\left.\h{3cm}{+\frac{g_{12}}{2\pi l_{12}^2} \op{\Psi}_1^\dagger\op{\Psi}_2^\dagger\op{\Psi}_1\op{\Psi}_2}\right],
\label{QF-Hamiltonian}
\end{eqnarray}
where  $g_\alpha=4\pi\hbar^2{\rm a}_\alpha/m_\alpha$ is the 3D interaction strength defined by the $s$-wave scattering length ${\rm a}_\alpha$, with $\alpha \in\{1,2,12\}$;  and $l_i$ are the harmonic oscillator lengths for the transverse confinement for the two species. The potential is taken to be a lattice with species-selective strength but with common period $2\pi/q$, with $q$ a natural number, and rotation rate $\Omega$:
\begin{eqnarray}
U_i(s,t)&=&\hbar u_{xi} \cos\left[2q ({\textstyle\frac{s}{R}} - \Omega_i t) \right]\nonumber\\&&+ \hbar u_{yi} \sin\left[2q ({\textstyle\frac{s}{R}} - \Omega_i t) \right].
\end{eqnarray}
We have allowed for two components of the lattice, one that is symmetric (denoted by $x$) and the other anti-symmetric (denoted by $y$) about the coordinate origin; and define the superpositions $u_{i\pm}=\frac{1}{2}(u_{xi}\pm iu_{yi})$. The explicit time-dependence in the Hamiltonian is removed by switching to a frame rotating with the lattice, which adds an angular momentum term.   The stationary modes $\frac{1}{\sqrt{2\pi R}}e^{in(s/R)}$ in the ring in the absence of the interaction have eigenenergies $\hbar\omega_n=\frac{\hbar^2 n^2}{2mR^2}$.  For small rings and low density, it was shown \cite{Opatrny-Das-two-species} that to good approximation we can confine the relevant dynamics to two circulating modes for each species $n\pm q$, commensurate with the lattice. We label the clockwise and counterclockwise mode amplitudes of each species, $\hat{a}_j, \hat{b}_j$, which satisfy the bosonic commutator rules. \emph{We will take the major radius $R$ as the length unit, the energy of the lowest mode $\hbar\omega_1=\frac{\hbar^2}{2mR^2}$ as the energy unit, and the associated frequency $\omega_1$ as the frequency unit}.

We cast the two-mode model in terms of collective spin operators:
\begin{eqnarray}\label{J-def}
    \hat{J}_{xi}=\frac{1}{2}\left(\hat{a}_i^\dagger\hat{b}_i + \hat{a}_i\hat{b}_i^\dagger\right), \nonumber \\
    \hat{J}_{yi}=\frac{1}{2i}\left(\hat{a}_i^\dagger\hat{b}_i - \hat{a}_i\hat{b}_i^\dagger\right), \nonumber \\
    \hat{J}_{zi}=\frac{1}{2}\left(\hat{a}_i^\dagger\hat{a}_i - \hat{b}_i^\dagger\hat{b}_i\right),
\end{eqnarray}
We introduce effective 1D interaction strengths $\chi_\alpha=\frac{g_\alpha}{4\hbar\pi^2 l^2R}$, and it was shown that for relevant physical systems can be well approximated by taking them to be equal $\chi_1=\chi_2=\chi_{12}=\chi$ \cite{Opatrny-Das-two-species}. Setting $\hbar =1$, the Hamiltonian can finally be expressed as a sum of linear and quadratic terms $\hat{H}=\hat{H}_L+\chi\hat{H}_Q$ as
\begin{eqnarray}
\op{H}_L&=&\sum_{i=1,2} \left[-2 q\Omega  \op{J}_{zi}+ u_{xi} \op{J}_{xi} +  u_{yi} \op{J}_{yi}\right]\n\\
\op{H}_Q &=&  \chi\left(\op{J}_{x1}+\op{J}_{x2}\right)^2 + \chi \left(\op{J}_{y1}+\op{J}_{y2}\right)^2
    \label{HQ_def}
\end{eqnarray}
neglecting  constants which do not affect the overall dynamics of the system. The linear portion corresponds to a rotations of the Bloch spheres; the quadratic portion changes the shape of the states, which leads to two-mode squeezing and generates entanglement between the two species.

\begin{figure}[t]
\centering
%\vspace{-0.2\linewidth}
\includegraphics[width=\columnwidth]{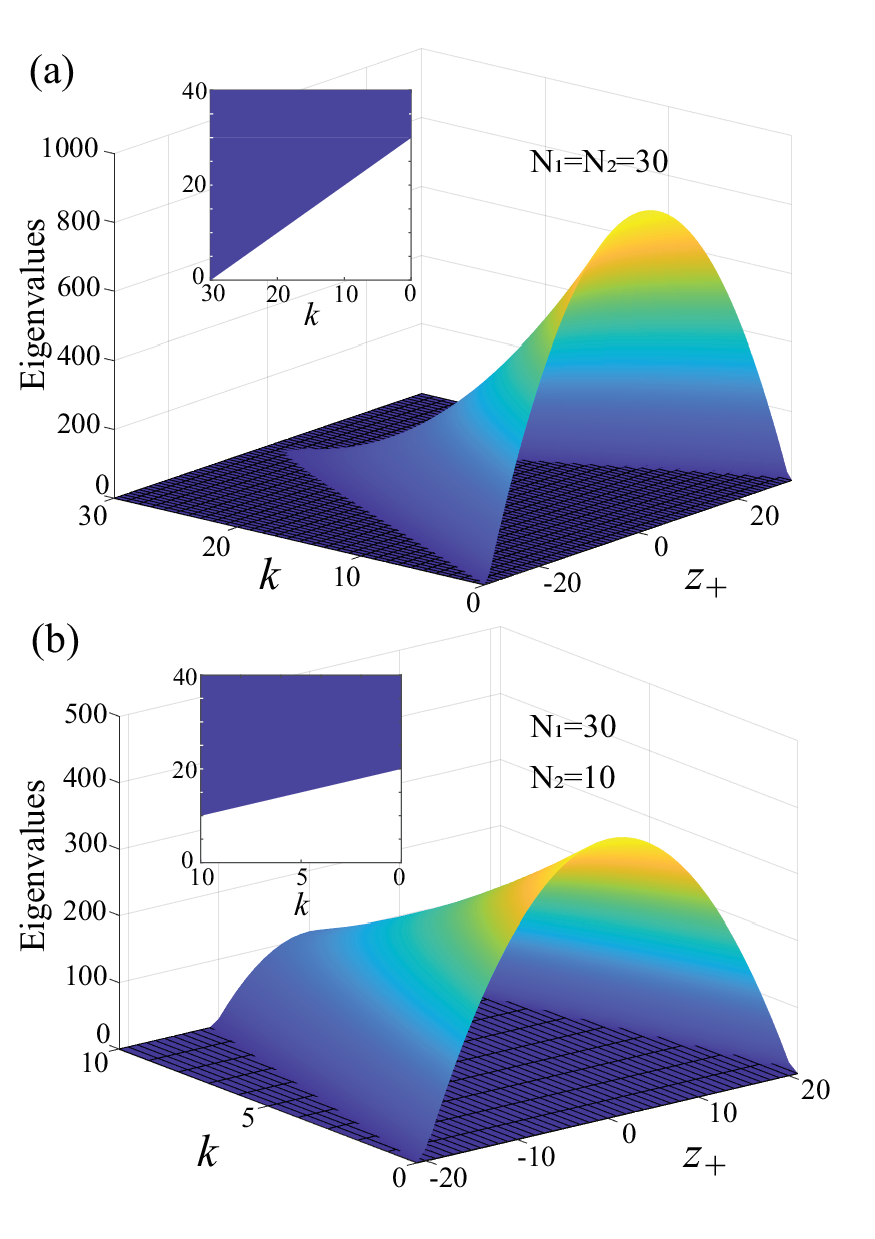}
\caption{Eigenvalues as a function of $z_+$ and dot-product index $k$. (a) Spectrum for $N_1=N_2=30$. When the number of atoms in each species are equal, the spectrum gradually decreases to a point with increasing $k$. The base of the distribution is a triangle, following the line $k=\frac{1}{2}N-|z_+|$. (b) Spectrum for $N_1=30$ and $N_2=10$. When the two numbers are unequal, there is an abrupt drop-off when $k$ is equal to the smaller of the two numbers. The base of this distribution is a trapezoid, or a triangle with its top cut off. The angled sides follow the $k=\frac{1}{2}N-|z_+|$ line, and the tip is cut off at $k_{max}={\rm min}(N_1,N_2)$. The insets show that the bottom of these surfaces are \emph{inclined} lines with the ground state corresponding to $k=k_{max}$.}
\label{eigenvalues}
\end{figure}

\section{Spectra of Linear and Quadratic Hamiltonians}
The linear Hamiltonian has the form of the Zeeman Hamiltonian \cite{Cohen_Tannoudji}, which describes the impact of a magnetic field on particles with non-zero spin. Thus $H_L$ can be interpreted as a magnetic field applied along some angle in 3D space as specified by the relative weights on each $\hat{J}_i$. To find the energy spectrum, one may take the angle at which the magnetic field is applied to be in the $\hat{J}_{\zeta}$ direction, with a magnitude given by the norm of all of the coefficients. The linear Hamiltonian can therefore be written as
\begin{eqnarray}\label{linear_Ham}
    H_L = \sum_{i=1,2}\sqrt{u_{xi}^2+u_{yi}^2+4q^2\Omega^2} \hat{J_{\zeta i}},
\end{eqnarray}
where $\zeta$ can be understood as the direction of the applied external magnetic field, and the spectrum is subsequently given by
\begin{eqnarray}
    E_{\zeta_1,\zeta_2} = \sum_{i=1,2} \gamma_i \zeta_i,
    \label{linear_spectrum}
\end{eqnarray}
where $\zeta_i \in \{-\frac{N_i}{2},-\frac{N_i}{2}+1,...,\frac{N_i}{2}\}$ are the eigenvalues of the respective $\hat{J}_{\zeta i}$, and $\gamma_i = \sqrt{u_{xi}^2+u_{yi}^2+4q^2\Omega^2}$ are the norm of all of the potential terms.

In order to determine the spectrum for the quadratic Hamiltonian $H_Q$, we need to determine its CSCO. For each individual species separately, $i=1,2$, in the collective spin picture we know that there are two distinct commuting observables besides the Hamiltonian, $\hat{J}_i^2$ and $\hat{J}_{zi}$. This indicates that for the system of two species together we should expect four such observables. It can be easily shown that $\hat{J}_{zi}$ do not commute with the Hamiltonian. But it was established that the three operators $\hat{J}_1^2$, $\hat{J}_2^2$, and $\hat{J}_{z1}+\hat{J}_{z2}$ commute with $H_Q$; and an incomplete picture was presented in terms of these operators in our prior work \cite{Opatrny-Das-two-species}. A complete characterization of the spectrum requires another quantum number that corresponds to a fourth operator that commutes with the Hamiltonian as well as the other three operators. We have now determined that the scalar product $\hat{J}_1\cdot\hat{J}_2$ completes the set of commuting observables for the quadratic Hamiltonian.  Thus for $H_Q$ the CSCO comprises of $\{H_Q, \hat{J}_1^2$, $\hat{J}_2^2$,$\hat{J}_{z1}+\hat{J}_{z2},\hat{J}_1\cdot\hat{J}_2\}$.

While the eigenvalues of the first three operators are well known in the context of collective spin operators, those of the fourth operator requires some calculation. For fixed values $N_1, N_2$ of the number of particles in each species, we found the eigenvalues of this commuting observable $\hat{J}_1\cdot\hat{J}_2$ to be given by
\begin{eqnarray}
E_k = {\textstyle \frac{N_1 N_2}{4}-\frac{1}{2}k(N_1+N_2+1-k)},
\end{eqnarray}
where we introduce the quantum number $k\in\{0,1,\dots k_{max}\}$, and
\begin{eqnarray}
    k_{max}={\rm min}\left(\frac{1}{2}N-|z_+|,N_1,N_2\right).
    \label{kmax}
\end{eqnarray}
Here $N=N_1+N_2$ is the total number of atoms in the two-species system.

\begin{figure*}[t]
\centering
%\vspace{-0.2\linewidth}
\includegraphics[width=\textwidth]{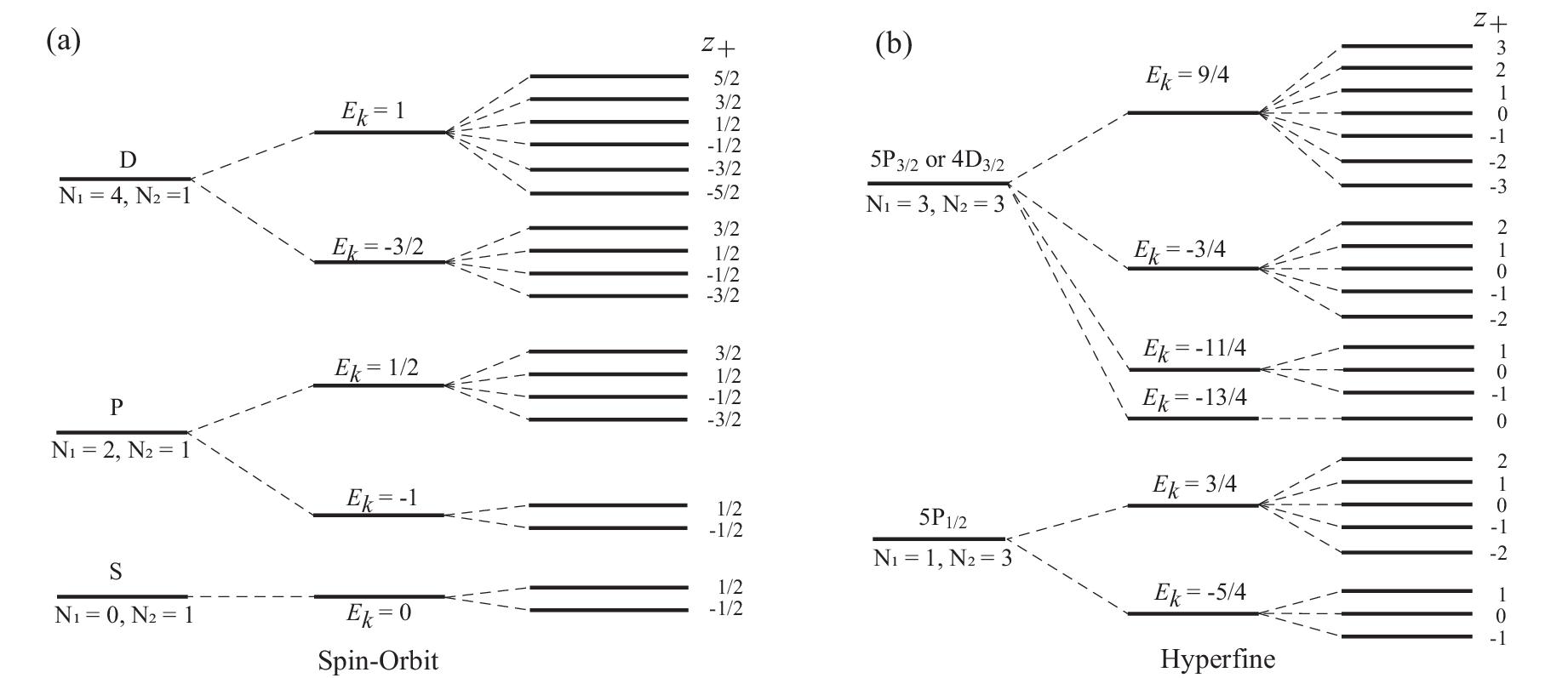}
\caption{The Hamiltonian in Eq.~(\ref{HQ_def}) can simulate both spin-orbit and hyperfine spectral features of electronic orbitals of atoms. (a) A single atom in one species and even number of atoms in the other recreates the spin-orbit coupling $\hat{L}\cdot\hat{S}$. (b) An odd number of atoms in one species to represent total electronic angular momentum including spin, $\hat{J}=\hat{L}+\hat{S}$, and either odd or even number in the other species to represent nuclear spin, can replicate the hyperfine structure $\hat{J}\cdot\hat{I}$; the case shown here simulates $^{87}$Rb with nuclear spin of $I=3/2$. In either case, the linear Hamiltonian with nonzero $\Omega$ and $u_{xi},u_{yi}\simeq 0$ lifts the degeneracy of $|z_+|\rightarrow z_+$. }
\label{hyperfine}
\end{figure*}

The quadratic Hamiltonian in Eq.~(\ref{HQ_def}) can now be written explicitly in terms of the four commuting observables
\begin{eqnarray}\label{HQ-CSCO}
\hat{H}_Q=\hat{J}_1^2 + \hat{J}_2^2 + 2\hat{J}_1\cdot\hat{J}_2 - (\hat{J}_{z1}+\hat{J}_{z2})^2 .
\end{eqnarray}
We denote the eigenvalues of the collective two-species operator $\hat{J}_{z1}+\hat{J}_{z2}$ by $z_+\in \{-\frac{N}{2},-\frac{N}{2}+1,...,\frac{N}{2}\}$ in obvious analogy of those for the species-specific counterparts in Eq.~(\ref{linear_spectrum}).
The spectrum of the quadratic Hamiltonian can now be characterized by the four quantum numbers $N_1,N_2,k, z_+$ corresponding to the four commuting observables. It is now straightforward to determine the eigenvalues of the Hamiltonian for well-defined particle number $N_1,N_2$ for the two species
\begin{eqnarray}
E_{k,z+} = {\textstyle \frac{N_1}{2}\left(\frac{N_1}{2}+1\right) + \frac{N_2}{2}\left(\frac{N_2}{2}+1\right) + \frac{N_1 N_2}{2} }\nonumber \\ -k\left( N_1+N_2+1-k\right) - z_+^2 .
\label{Hdot_eigvals}
\end{eqnarray}
There are $g_k=N+1-2k$ states with the same quantum number $k$, and there is a twofold degeneracy with respect to $\pm z_+$ except when $z_+=0$.

The representation of the spectrum as a function of $k$ and $|z_+|$ provides a more straightforward picture of the energy spectrum. The spectrum can be represented as having a triangular base with sides following $k=\frac{1}{2}N-|z_+|$. The eigenvalues follow a downward sloping surface, as shown in Fig.~\ref{eigenvalues}(a). When $N_1\neq N_2$, the tip of the base is cut off at the minimum number of particles, shown in Fig.~\ref{eigenvalues}(b). Equation~(\ref{Hdot_eigvals}) implies that all the states are at least doubly degenerate with states with $\pm z_+$ having the same energies. The exception to this is when $z_+=0$. When both species have the same number of particles, the ground state has $z_+=0$, and as the inset in Fig.~\ref{eigenvalues}(a) illustrates the ground is non-degenerate. On the other hand, when the numbers are not the same, the ground state does not have $z_+=0$ and is doubly degenerate. Due to the scale of the eigenvalues plotted in the surface plots in Fig.~\ref{eigenvalues}, it is not immediately clear that the bottom edges are inclined lines, which we highlight in the insets.

\begin{figure*}[t]
\centering
%\vspace{-0.2\linewidth}
\includegraphics[width=\textwidth]{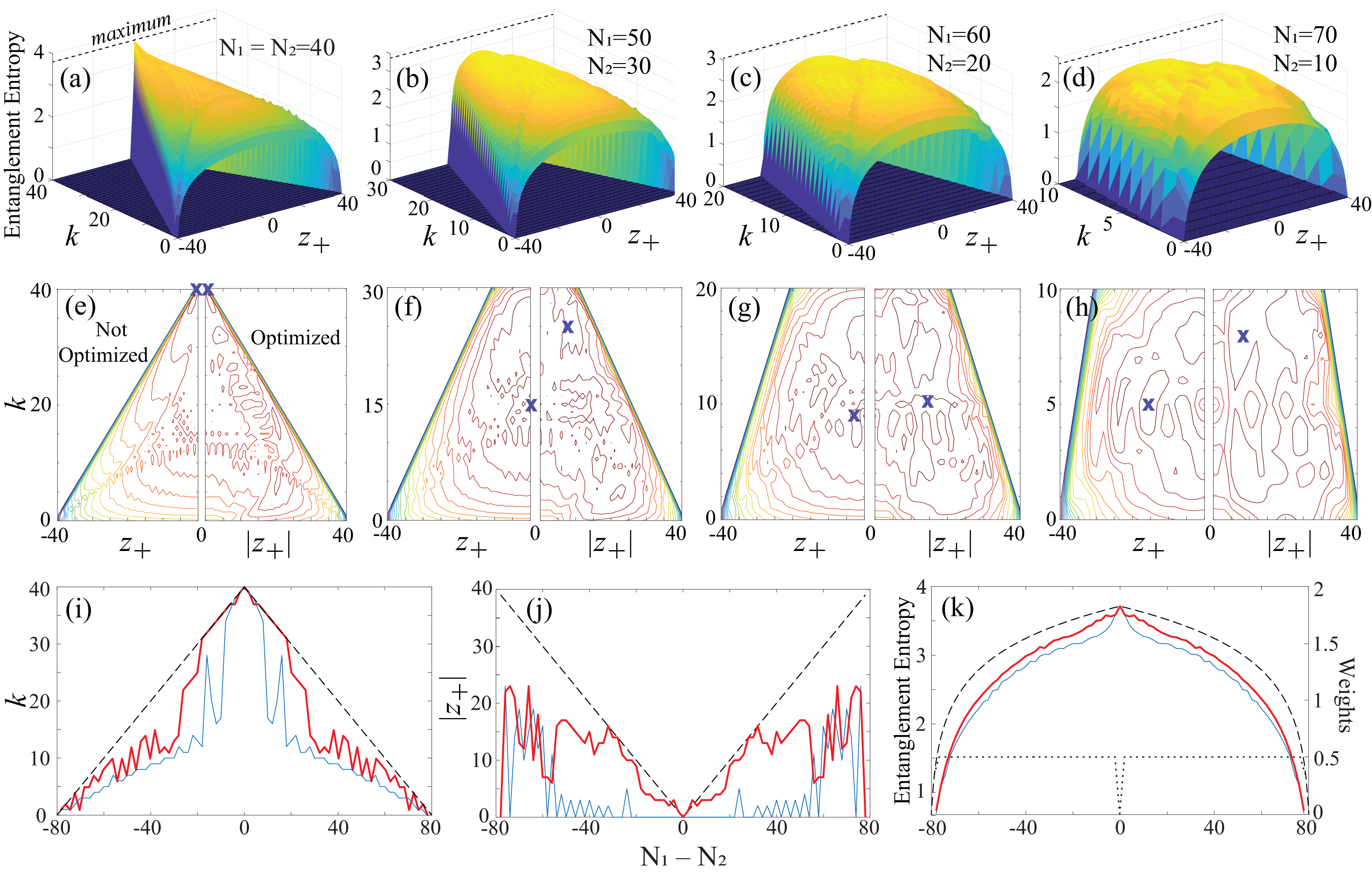}
\caption{Entanglement of eigenstates of the quadratic Hamiltonian as a function of dot product index $k$ and value of $z_+$ for a fixed total number of particles $N=80$. The surface and contour plots, respectively, of the EE for increasing imbalance: (a,e) $N_1=N_2=40$; (b,f) $N_1=50$ and $N_2=30$; (c,g) $N_1=60$ and $N_2=40$; and (d,h) $N_1=70$ and $N_2=10$. On the surface plots, the maximum possible EE is marked with a dashed line. On the contour plots, states with highest entropy are marked with crosses; the left half of (e-h) use eigenstates of $H_Q$ with fixed $k$ and $z_+$, and the right half a linear combination of eigenstates with the same $k$ and $\pm z_+$ that maximizes entanglement. The values of $k$ and $z_+$ at maximum EE are shown on (i) and (j), respectively, with the thin blue line for the un-optimized scheme and the thick red line for the optimized scheme. The dashed lines in (i,j) represent the values of $k$ and $|z_+|$ for the ground state. (k) The corresponding maximum EE values, un-optimized (thin blue) and optimized (thick red) are plotted along with the maximum possible EE (dashed line) all along the \emph{left axis}; the optimal weights of the two eigenstates are plotted in dotted line along the  \emph{right axis}.}
\label{entropy}
\end{figure*}

\section{Simulating Spin-Spin interactions}

The quadratic Hamiltonian as written in Eq.~(\ref{HQ-CSCO}) has the unusual characteristic that it comprises of a linear combination of all the operators of its CSCO, noting that any power of commuting observables will still commute with each other and the Hamiltonian. That allows flexibility in the choice of the values for each term in the Hamiltonian. Specifically, we can fix the first two terms $\hat{J}_1^2 + \hat{J}_2^2$ by fixing the number of particles. Within that subspace, we can consider the variation of the remaining two terms. Up to a constant factor, the next term is the product $\hat{J}_1\cdot\hat{J}_2$, which represents interaction of two independent angular momenta.  This can clearly be adapted to model the full range of such interactions which occur typically in subatomic regimes of electronic and nuclear states. The difference is that in the ring system the behavior of such interactions can be studied in the context of external translational quantum states of atoms at a substantially larger spatial scale and proportionately slower time scale. Furthermore, the engineered nature of the system allows such interactions to be controlled in a way not possible in their naturally occurring subatomic counterparts. We illustrate the versatility by considering two specific cases of strong interest across several areas of physics: (1) Spin-orbit coupling which gives rise to the fine-structure features in atomic spectra and (2) hyperfine splitting of spectral lines.

In the collective spin picture in the ring, it is clear that each atom can be treated as representing spin $1/2$ with its two counter-propagating modes ($\hat{a}, \hat{b}$) representing spin up and spin down respectively. This is reflected in the collective spin operators in Eq.~(\ref{J-def}) and the associated eigenvalues described in the previous section. If we fix the particles to be such that one species has even number of particles, say $N_1$ and the other species comprises of only one particle, $N_2=1$, that will be equivalent to having species one with integer spin given by $L=N_1/2$ and species two with spin $S=1/2$.  Then $\hat{J}_1\cdot\hat{J}_2=\hat{L}\cdot\hat{S}$ represents the coupling of the electron spin with the orbital angular momentum, that is the spin-orbit coupling which gives rise to the fine-structure in atomic spectra. The energy splitting diagram is shown in the middle column of Fig.~\ref{hyperfine}(a), with the leftmost column representing the unsplit $S,P,$ and $D$ electronic orbitals of an atom.

If, on the other hand, we take species one to have an odd number of particles, so that $N_1$ is odd, then that can represent the sum $\hat{J}=\hat{L}+\hat{S}$, corresponding to the total angular momentum of an orbiting electron.  The other species can be any non-negative integer such that $N_2/2$ corresponds to a nuclear spin $\hat{I}$. In this scenario,  the product term $\hat{J}_1\cdot\hat{J}_2=\hat{J}\cdot\hat{I}$, would simulate a hyperfine splitting of spectral lines. This is illustrated for the case $^{87}$Rb in Fig.~\ref{hyperfine}(b) which has a nuclear spin of $I=3/2$, with the leftmost column representing some of the corresponding fine structure lines.

In both cases described above, to present further splitting due to the presence of an external magnetic field, which would differentiate states with different values of $J_z$, we need to use the fourth term $(\hat{J}_{z1}+\hat{J}_{z2})^2 $ in the Hamiltonian Eq.~(\ref{HQ-CSCO}). Unlike in the case of a real magnetic field, this term being squared cannot lift the degeneracy between states with $\pm|z_+|$. But the degeneracy may be lifted by introducing non-zero rotation of the ring represented by the first term in the linear Hamiltonian $\hat{H}_L$. Note, with $u_{xi},u_{yi}\simeq 0$, $\zeta_i\simeq z_i$ upto a scale factor. This is interesting due to the similarity between rotation and magnetic fields that is ubiquitous in quantum mechanics. The fully split spectra including the effect of this term is shown in the rightmost columns of Figs.~\ref{hyperfine}(a,b).

\section{Entanglement Entropy}
Beyond realizing them in this macroscopic setup, the actual applications of the spin-spin interaction, such as in spintronics or quantum information, would require a knowledge of the degree of entanglement possible in such systems. We can measure this for a pure state by computing the von Neumann entropy using the reduced density matrix $\rho_1 = \text{Tr}_2(\rho)$ or $\rho_2 = \text{Tr}_1(\rho)$, where $\rho=\ket{\Psi}\bra{\Psi}$ is the density matrix. The von Neumann entropy is
\bn
S(\rho_1) = \text{Tr}[\rho_1\ln \rho_1] = -\sum_i [\epsilon_i \ln(\epsilon_i)],
\en
where we assume $\rho_1$ is diagonalizable and $\epsilon_i$ are its eigenvalues. The entropy does not depend on the choice of reduced density matrix, so that $S(\rho_1) = S(\rho_2)$.

With our knowledge of the CSCO, we fix the number of particles in each species, and compute and plot the entanglement entropy (EE) as a function of the quantum numbers  $k$ and $z_+$  associated with the remaining two commuting observables. In Fig.~\ref{entropy}, we plot the EE for a varying degree of imbalance of the particle number of the two species $N_1-N_2$, with the total number fixed at $N_1+N_2=80$.  The top row, panels (a-d), displays the EE as surface plots, with the maximum possible entropy $ S_{max}=\ln|1+\min(N_1,N_2)|$ indicated. The plots are bilaterally symmetric about the $z_+$ axis due to the two-fold degeneracy mentioned earlier.

The second row, panels (e-f), displays the EE as contour plots, with left side of each being the counterpart of the surface plots. The right side uses an optimized linear combination of the pair of normalized degenerate eigenstates $c_+\phi(k,z+)+c_-\phi(k,-z+)$ with $c_\pm$ being a complex numbers chosen to maximize the EE, and contour plots have $|z_+|$ along the horizontal axis. The maximum EE is generally increased by the optimization process as we might expect, but it is interesting to note that there is a shift in the $k, |z_+|$ values where the maximum occurs, as indicated by the shifting of the crosses between the left and the right panels.

In the bottom row of Fig.~\ref{entropy}, we compare the optimized with the non-optimized EE, plotting the $k$ values in panel (i) and then the $|z_+|$ values in panel (j), that correspond to the maximum EE for each case. The non-optimized ones uses one of the degenerate ground states $\phi(k,-z_+)$, the same as if we used $\phi(k,z_+)$, while the optimized case uses their linear combination as previously described. The dashed lines in both plots mark the value of $k$ and $|z_+|$ for the ground state; those coincide with $k=k_{max}$, the maximum possible values as indicated in Eq.~(\ref{kmax}), but not the maximum for $|z_+|$ which is always at $N/2$.   Panels (i) and (j) show that the EE is minimal for $k=0$ and $|z_+|=\textstyle \frac{N}{2}$, and generally increases with increasing $k$ and decreasing $|z_+|$, though there is a more complex internal pattern. For $N_1=N_2$, the highest EE occurs at $z_+=0$ and $k=k_{max}$, and the lowest at $k=0$ and $z_+=\pm \textstyle \frac{N}{2}$. The key message of panels (i) and (j) of Fig.~\ref{entropy} is that the maximum EE occurs at the ground state (marked by the dashed line) for $N_1\approx N_2$ and, when the optimized ground state is used, continues for a moderate range of increasing asymmetry between $N_1$ and $N_2$.

Fig.~\ref{entropy}(k) plots the maxiumum EE achieved with the optimized states versus the non-optimized states. We see the optimized states increase the EE across all degrees of asymmetry, moving closer to the maximum possible EE as marked by the dashed line. We found the optimal mixture was almost always an equal mix $|c_\pm|^2=0.5$ as indicated by the dotted line, and that the relative phase of coefficients $c_\pm$ had no impact on the EE. Therefore, to prepare states with the maximum entanglement, one may prepare the ground state for roughly equal number of particles, with a wider range about $N_1=N_2$ permitted with eigenstate optimization.

\begin{figure*}[t]
\centering
%\vspace{-0.2\linewidth}
\includegraphics[width=\textwidth]{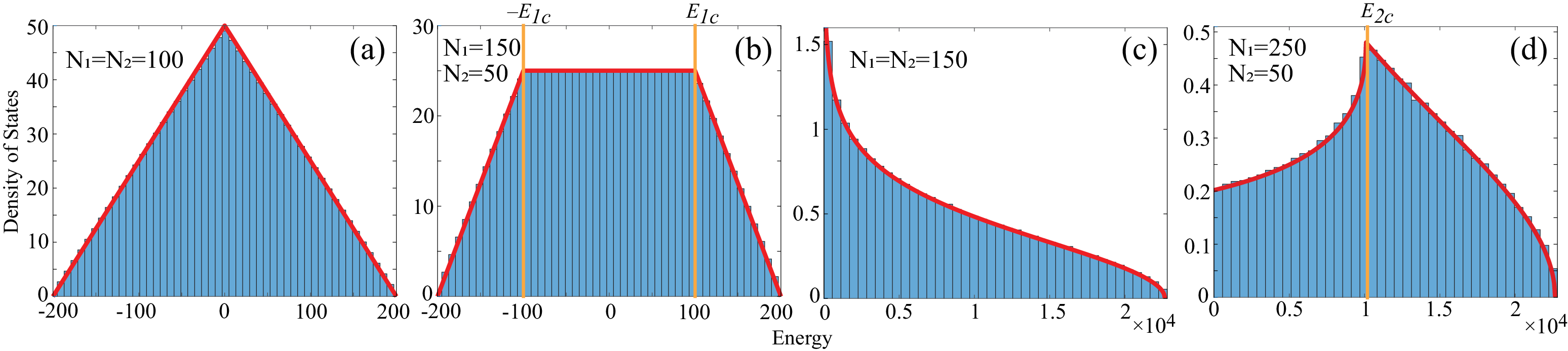}
\caption{Histograms of the density of states (DOS), calculated directly from the Hamiltonian, with the analytical DOS overlaid (red solid line), for (a) $H_L$ with $N_1=N_2=100$ and $\gamma_1=\gamma_2=2$; (b) $H_L$ with $N_1=150$, $N_2=50$, and $\gamma_1=\gamma_2=2$;  (c) $H_Q$ with $N_1=N_2=150$; and (d) $H_Q$ with $N_1=250$, $N_2=50$.  The critical energies $E_{1c}$ and $E_{2c}$ that mark the transition to a dependence only on the total particle number are marked by light orange vertical lines. A constant factor was introduced to make the analytical and numerical values have the same scale, the matching shape being the relevant point. }
\label{state_dens}
\end{figure*}

\section{Density of States}
In order to complete the picture we now describe the density of states (DOS) of this system.  In our previous paper \cite{Opatrny-Das-two-species}, we remarked on the changing characteristic of the DOS as the system is transformed from being in the purely linear regime to the purely quadratic. That analysis was done numerically, but now with the CSCO at hand, we can compute the DOS analytically in the two limiting cases. We may directly calculate the DOS as
\bn
    \frac{dn}{dE} = \int \frac{d\sigma}{|\nabla E|},
    \label{grad_dens}
\en
where $n$ is the number of states, $|\nabla E|$ is the norm of the gradient with respect to the relevant set of complete variables designated by $\sigma$, and the line integral is taken over the lines of equal energy. Here we will assume that  the  number of particles in the two species, $N_1$ and $N_2$, are fixed.

We first consider the linear Hamiltonian $H_L$ in Eq.~(\ref{linear_Ham}), for which $\sigma\equiv (\zeta_1,\zeta_2)$ and the energy is given by Eq.~(\ref{linear_spectrum}), so that
\bn
    |\nabla E| = \sqrt{\gamma_1^2+\gamma_2^2}, \h{1cm}
    d\sigma = \frac{d\zeta_2}{\gamma_1}\sqrt{\gamma_1^2+\gamma_2^2}.
\en
With the resulting cancellations, and by expressing the limits of $\zeta_2$ as a function of the energy $E$ by rearranging Eq.~(\ref{linear_spectrum}), the integral is easily evaluated to yield the DOS for the linear Hamiltonian:
\begin{eqnarray}
    \left.\frac{dn}{dE}\right|_L = \frac{1}{\gamma_1}\left[\min\left(\frac{1}{\gamma_2}(E+\gamma_1\tfrac{N_1}{2}),\frac{N_2}{2}\right) \hspace{1cm}\right. \nonumber \\
    \left. - \max\left(\frac{1}{\gamma_2}(E-\gamma_1\tfrac{N_1}{2}),-\frac{N_2}{2}\right) \right].
\end{eqnarray}
We plot this in Figs.~\ref{state_dens} in thick red lines and compare with numerically evaluated histograms for the corresponding distribution, for equal number of particles $N_1=N_2=100$ in panel (a) and for different numbers $N_1=150, N_2=50$ in panel (b). There is clear agreement of the analytically computed trace with the numerical results. The DOS for $H_L$ takes one of two shapes: triangular or trapezoidal. The triangular shape is a special case which arises when $\gamma_1 N_1=\gamma_2 N_2$, as seen in Fig.~\ref{state_dens}(a). The trapezoidal shape occurs when $\gamma_1 N_1\neq \gamma_2 N_2$.

Next we turn to the more interesting case of the quadratic Hamiltonian $H_Q$ in Eq.~(\ref{HQ-CSCO}). Now the integration variables are $\sigma=(k,z_+)$, and the energy is given by Eq. (\ref{Hdot_eigvals}).  By rearranging the expression for the energy,
\bn
    E +{\textstyle\frac{1}{4}} = \left(k-\frac{N_1+N_2+1}{2}\right)^2 - z_+^2
\en
we can parameterize the integration variables as
\bn
    z_+ &=& \sqrt{E+{\textstyle\frac{1}{4}}}\ \text{sinh}\xi, \nonumber \\
    k &=& {\textstyle\frac{1}{2}}(N_1+N_2+1)- \sqrt{E+{\textstyle\frac{1}{4}}}\ \text{cosh}\xi.
\en
We first consider the special case of an equal number of particles in the two species, $N_1=N_2$, in which case the parameter $\xi\in [-\xi_{max}, \xi_{max}]$, where
\bn
    \xi_{max} = \text{acosh}\frac{N_1+N_2+1}{\sqrt{4E+1}}.
\en
The energy gradient and the line element in parameter space then become
\bn
 |\nabla E| &=& \sqrt{4E+1}\sqrt{\text{cosh}^2\xi+\text{sinh}^2\xi}\n\\
    d\sigma &=& \sqrt{E+{\textstyle\frac{1}{4}}}\sqrt{\text{cosh}^2\xi+\text{sinh}^2\xi} d\xi.
\en
The integral in Eq.~(\ref{grad_dens}), being along lines of constant $E$, reduces to an integral over $\xi$, and we compute the DOS for $N_1=N_2=N/2$ to be
\bn
    \frac{dn}{dE} = \text{acosh} \frac{N+1}{\sqrt{4E+1}}.
    \label{denseqn_equal}
\en
In the general case, when the number of particles of the two species are unequal $N_1\neq N_2$ we know that the spectrum takes the same shape as if the two were equal, except that spectrum above $\text{min}(N_1,N_2)$ is excised, as shown in Fig.~\ref{eigenvalues}. Therefore, another term is necessary to subtract the missing piece from the DOS derived above for equal numbers of particles in both species. The equation can be parameterized in exactly the same way as for equal number of particles, resulting in an identical integral, but the limits of $\xi$ will change to account for the states with values of $k$ above $\text{min}(N_1,N_2)$:
\bn
    \xi_{max} = \text{acosh}\frac{|N_1-N_2|+1}{\sqrt{4E+1}}.
\en
Using this, the general DOS for the quadratic Hamiltonian is evaluated to be
\bn
    \frac{dn}{dE} = {\rm Re}\left\{\text{acosh} \frac{N+1}{\sqrt{4E+1}} - \text{acosh} \frac{|N_1-N_2|+1}{\sqrt{4E+1}}\right\}.\h{5mm}
    \label{denseqn}
\en
In the case of equal number of particles, $N_1=N_2$, the second term above becomes imaginary due to the properties of inverse hyperbolic functions, and the expression reduces to the first term which matches Eq.~(\ref{denseqn_equal}).

In Fig.~\ref{state_dens} we overlay the DOS calculated from this expression in thick red line  on a histogram for the distribution calculated numerically from the values and the number of eigenvalues of the quadratic Hamiltonian, first for equal number of particles $N_1=N_2=150$ in panel (c); and then for unequal number of particles $N_1=250, N_2=50$ in panel (d). The analytical results match the numerical results better as particle number increases, due to the assumption of a continuum limit in the analytical calculations. The general pattern we observe is that for equal number of particles, the DOS is skewed to lower energies, peaking at the ground state.  As the imbalance $|N_1-N_2|$ increases, the second term becomes prominent and gradually skews the DOS towards maximum energy, and also there are fewer states available. In the limit of extreme imbalance when almost all the particles belong to one species, the DOS is concentrated at or near the maximum energy.

There is a point at which the second term in Eq.~(\ref{denseqn}) becomes purely imaginary due to the behavior of inverse hyperbolic functions. This transition occurs when the total energy \emph{exceeds}
\bn
    E_{2c} = \frac{|N_1-N_2|}{2}\left(\frac{|N_1-N_2|}{2}+1\right).
\en
At energies above this value, the density of states is given by the real value of the expression in Eq. (\ref{denseqn}), which is equivalent to Eq. (\ref{denseqn_equal}).  This means that above a certain energy,  the DOS for the quadratic Hamiltonian is insensitive to how the particles are partitioned between the two species, as in the specific values of $N_1$ and $N_2$, and depends only on the total number of particles. This is what we would expect from the shape of the spectra in Fig.~\ref{eigenvalues}, since the spectrum for unequal number of particles becomes identical with that of equal number of particles, for the same total number, above a certain energy.

This behavior has a counterpart in the linear Hamiltonian, which retains the same structure of the density of states for a fixed total number of particles, except that some of the states with energy around zero are not available when there is an asymmetry in either the number of particles or the norms of the potentials, $\gamma_1$ and $\gamma_2$. This results in the trapezoidal shape, as in Fig.~\ref{state_dens}(b), effectively a truncated triangle, with the missing part of the triangle corresponding to the unavailable states. However in the case of the linear Hamiltonian, $H_L$, there is a symmetry about the states with zero energy due its dependence on $z_+$ rather than on $z_+^2$. This means that the DOS is identical for a fixed total number of particles, regardless of the partition between the two species or values of $\gamma_1$ and $\gamma_2$, for energies $|E|>E_{1c}$ where the critical energy is
\bn
 E_{1c} = {\textstyle\frac{1}{2}}|\gamma_1 N_1 - \gamma_2 N_2|.
\en
which marks the boundaries of the narrower upper edge of the trapezoid as seen in Fig.~\ref{state_dens}(b).

\section{Outlook and Conclusions}

We have shown that two species of BEC in a ring trap can be adapted to simulate interactions of spin and angular momentum, specifically spin-orbit coupling and hyperfine interaction. The Hamiltonian describing the system has a linear and a quadratic component. The linear part can be controlled using the strength and depth of an azimuthal lattice and can be used to initialize the quantum states. The quadratic part depends on interparticle interactions which can be manipulated with Feshbach resonances \cite{Tiesinga-RMP} or changing the effective 1D density by adjusting the transverse confinement.

The spectrum for both the linear and quadratic components were determined analytically.  The linear part serves as a model for a Zeeman Hamiltonian. We found the CSCO for the quadratic Hamiltonian and showed that it has the unusual feature of being a linear combination of the four CSCO variables, with one commuting observable appearing quadratically. We determined general analytical expressions for the spectrum for both linear and quadratic components, as well as the corresponding density of states that agreed well with numerical computed distributions. Unequal particle numbers present different qualitative spectra and DOS compared to having equal particle numbers. However, both the linear and the quadratic Hamiltonians feature a significant range of total energy wherein the DOS state depends only on the total number of particles in the system and not on how they are partitioned between the two species.

We computed the entanglement entropy of the two species system, relevant for diverse applications of interacting spin systems. For equal number of particles in the two species, the entanglement is always maximum for the ground state of the system, whereas for unequal particle numbers, the maximum does not occur at the ground state. Optimizing the linear combination of degenerate states was shown to lead to higher EE, with maximum EE occurring closer to the ground state for a broader range of interspecies population imbalance.

The results of this paper can be tested and implemented by merging existing experimental capabilities with ultracold atoms in toroidal traps \cite{Phillips_Campbell_superfluid_2013,Hadzibabic-2012, von_Klitzing_BEC_accelerator} with methods for creating ring shaped lattice potentials \cite{Padgett}.  As shown here, the system can then serve as a quantum simulator for spin-based physics, that includes spin-spin interactions like spin-orbit and hyperfine coupling. Compared to the dynamics of intrinsic spins, this system offers much larger size and slower time scales that can be assets in some applications and for probing fundamental physics. These can benefit greatly from the numerous physical parameters like lattice depths, density, interaction strengths that can be controlled with high precision in ultracold atomic systems. With two species of interacting collective spins, the system can be used to study and utilize a multitude of correlation effects that arise from the nonlocal nature of quantum systems, but here, with the advantage of macroscopic scales.

\begin{acknowledgments} This work was supported by the Czech Science
Foundation Grant No. 20-27994S for T. Opatrn\'y and by the NSF under Grant No. PHY-2011767 and PHY-2309025 for Kunal K. Das.  \end{acknowledgments}

\end{document}